\documentclass{PoS}

\newcommand{\beq}{\begin{equation}}
\newcommand{\eeq}{\end{equation}}
\newcommand{\bea}{\begin{eqnarray}}
\newcommand{\eea}{\end{eqnarray}}

\newcommand{\mr}{\mathrm}
\newcommand{\hs}{\hspace*}

\title{Weak boson scattering at the Large Hadron Collider}

\ShortTitle{Weak boson scattering at the LHC}

\author{\speaker{B.~J\"ager}\\
        Institut f\"ur Theoretische Physik und Astrophysik, 
	Universit\"at W\"urzburg, 97074 W\"urzburg, Germany
	}

\author{G.~Bozzi
\\
Dipartimento di Fisica, Universit\`a di Milano and INFN,
Sezione di Milano, Via Celoria 16, 20133 Milano, Italy
}

\author{C.~Englert, D.~Zeppenfeld
\\
Institute for Theoretical Physics, Karlsruhe Institute of Technology, 76128 Karlsruhe, Germany 
}

\author{C.~Oleari
\\
Universit\`a di Milano-Bicocca and INFN, Sezione di Milano-Bicocca, 
Piazza della Scienza 3, 20126 Milan, Italy
}
\author{M.~Worek
\\
Fachbereich C Physik, Bergische Universit\"at Wuppertal, 
42097 Wuppertal, Germany 
}

\abstract{
Weak boson scattering processes provide particularly promising means for gaining insight into the mechanism of electroweak symmetry breaking at hadron colliders.
Being very sensitive to interactions in the weak gauge boson sector, they will help to distinguish the Standard Model from various new physics scenarios such as extra-dimensional Higgsless models.  
To unambiguously identify signatures of new physics, precise predictions for experimentally accessible observables within realistic selection cuts are crucial, including next-to-leading order QCD corrections. 
Here, we review how flexible Monte-Carlo methods can be employed for precision analyses of weak boson scattering reactions within the Standard Model and beyond.
}

\FullConference{RADCOR 2009 - 9th International Symposium on Radiative Corrections (Applications of Quantum Field Theory to Phenomenology) ,\\
		 October 25 - 30 2009\\
		 Ascona, Switzerland}

\begin{document}
%
%%%%%%%%%%%%%%%%%%%%%%%%%%%%%%%%%%%%%%%%%%%%%%%%%%%%%%%%%%%%%%%%%%%
%
\section{Introduction}
Weak boson fusion (WBF) processes have been identified as a particularly promising class of reactions for gaining insight into the mechanism of electroweak symmetry breaking. Higgs production via WBF is considered as possible discovery mode for the iso-scalar, scalar resonance predicted by the Standard Model (SM). Once a SM-like Higgs boson has been found, WBF processes will help to determine its spin and CP properties and measure its couplings to gauge bosons and fermions. In many physics scenarios beyond the SM, electroweak symmetry breaking is realized by new interactions in the weak sector. 
Bulk-gauged extra-dimensional Randall-Sundrum models~\cite{Randall:1999ee} feature, for instance, 
infinite towers of new massive vector resonances, referred to as ``Kaluza-Klein excitations''. 
As shown in \cite{Csaki:2003dt,Cacciapaglia:2004rb}, one can arrive at models that implement electroweak symmetry breaking 
by appropriately chosen conditions on the boundaries of the static finite-sized Randall-Sundrum background for the 
gauge fields. Thereby any scalar is removed from the theory's spectrum, giving rise to an effective ``Higgsless model'' in four
dimensions. 
In WBF processes the signatures of such non-SM like scenarios should be pronounced and well-observable \cite{Birkedal:2004au,Englert:2008tn}, as a~priori large background processes can be tamed efficiently by the application of dedicated selection criteria. 

In order to unambiguously distinguish various signatures of new physics from the SM scenario, a precise, quantitative understanding of weak boson scattering reactions is essential, requiring the computation of next-to-leading order (NLO) QCD corrections to electroweak $VVjj$ production ($V$~denotes a $W^\pm$ or a $Z$ boson). Experimentally, very clean signatures are expected from the leptonic decay modes of the weak gauge bosons. 
Being implemented in a flexible parton-level Monte-Carlo program, the kinematic features of this class of reactions can be explored, allowing for the design of selection criteria that help to distinguish the WBF signal from various QCD backgrounds. 

In this contribution, we will review the NLO-QCD calculations that have been performed for weak boson scattering processes within the 
SM~\cite{Jager:2006zc} %,Jager:2006cp,Bozzi:2007ur,Jager:2009xx
and a representative model of new physics~\cite{Englert:2008wp}, 
taking leptonic decay correlations fully into account. In each case, NLO-QCD corrections to total cross sections are at the few-percent level and residual scale uncertainties of the NLO results are small. However, the shapes of some distributions change noticeably when going from LO to NLO. The application of dedicated selection cuts should allow for the separation of the WBF signal from various backgrounds~\cite{Englert:2008tn}.   

%%%%%%%%%%%%%%%%%%%%%%%%%%%%%%%%%%%%%%%%%%%%%%%%%%%%%%%%%%%%%%%%%%%
%
\section{Outline of the Calculation}
WBF production of a $4~\mathrm{leptons}+2~\mathrm{jets}$ final state in 
$pp$ collisions mainly proceeds via the scattering of two (anti-)quarks by $t$-channel exchange of a weak boson with subsequent emission of two vector bosons, which in turn decay leptonically. Non-resonant diagrams, where leptons are produced via weak interactions in the $t$ channel also have to be considered. 
Various interference effects and same-flavor annihilation contributions are negligible in the phase-space regions where WBF can be observed experimentally and therefore entirely disregarded (see, e.g., \cite{Bredenstein:2008tm} %,Ciccolini:2007ec
for explicit predictions for these contributions in the WBF $Hjj$ mode). 
The calculation of the relevant tree-level matrix elements is straightforward and can be accomplished numerically with the amplitude techniques of Refs.~\cite{Hagiwara:1985yu}. %,Hagiwara:1988pp

At NLO, real-emission and virtual corrections to the Born amplitude arise. Infrared singularities emerging in intermediate steps of the calculation are regularized in $d=4-2\epsilon$ dimensions and handled with the dipole subtraction formalism of Ref.~\cite{Catani:1996vz}. 
The real-emission contributions are obtained by attaching an extra gluon to the tree-level diagrams in all possible ways, giving rise to (anti-)quark initiated subprocesses with an additional gluon in the final states as well as contributions with a gluon in the initial state. 

The virtual corrections comprise the interference of one-loop diagrams with the Born amplitude. Due to the color-singlet nature of the $t$-channel weak boson 
exchange, only self-energy, triangle-, box-, and pentagon corrections to either the upper or the lower quark line have to be considered. The singularities of these 
contributions associated with infrared-divergent configurations are calculated analytically and canceled by respective poles in the integrated counter-terms of the 
dipole subtraction approach. The finite terms are evaluated numerically by the tensor reduction procedures of 
Refs.~\cite{Passarino:1978jh}. %,Denner:2002ii,Denner:2005nn
For details of the calculation, the reader is referred to Refs.~\cite{Jager:2006zc,Englert:2008wp}. %Jager:2006cp,Bozzi:2007ur,Jager:2009xx,
     
%%%%%%%%%%%%%%%%%%%%%%%%%%%%%%%%%%%%%%%%%%%%%%%%%%%%%%%%%%%%%%%%%%%
%
\section{Results}
The cross-section contributions discussed above for the various production modes have been implemented in a flexible parton-level Monte-Carlo program which allows the user to compute cross sections and kinematical distributions within the SM and a Warped Higgsless model for experimentally feasible selection cuts~\cite{Arnold:2008rz}. Here, a few representative results for WBF $W^+W^+jj$ and  $W^+Zjj$ production are shown. 

We use the CTEQ6M parton distributions with $\alpha_s(m_Z)=0.118$ at NLO and the CTEQ6L1 set at LO. We chose $m_W=80.423$~GeV, $m_Z=91.188$~GeV, and $G_F=1.166\times~10^{-5}/$~GeV$^2$ as electroweak input parameters. Thereof, $\alpha_\mathrm{QED}$ and $\sin^2\theta_W$ are computed via LO electroweak relations. Jets are reconstructed from final-state partons via the $k_T$ algorithm with resolution parameter $D=0.7$. Contributions from external $b$- and $t$-quarks are neglected and fermion masses are set to zero throughout. If not stated otherwise, we consider $pp$ collisions at a center-of-mass (c.m.s.) energy of $\sqrt{S}=14$~TeV. 
\begin{figure}[!t]
\begin{center}
  \includegraphics[width=0.6\textwidth,clip]{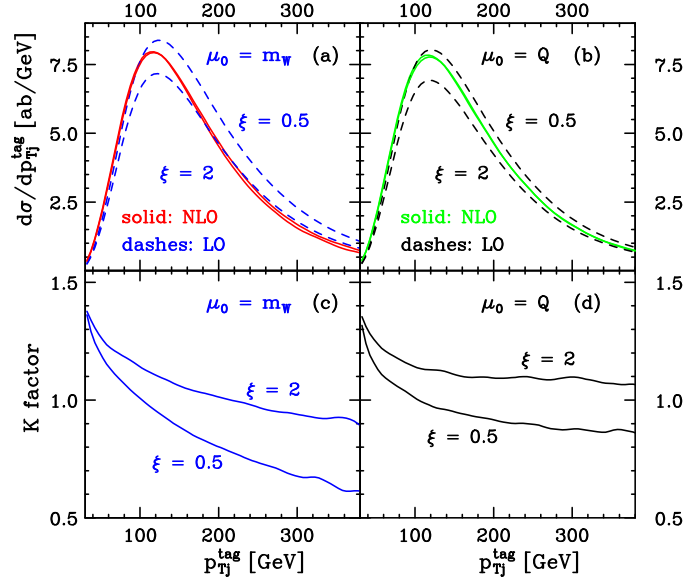}
  \caption{Transverse momentum distribution of the tagging jet with the highest $p_T$ in $pp\to e^+ \nu_e \mu^+\nu_{\mu} jj$ via WBF at LO (dashed lines) and NLO (solid lines) at the LHC for two different choices of $\mu_0$ [panels (a) and (b)] at LO (dashed) and NLO (solid). The corresponding $K$ factors are displayed for $\mu_0 = m_W$ in panel~(c) and for  $\mu_0 = Q$ in panel (d).
  }
  \label{fig:ptj}
\end{center}
\end{figure}
In order to clearly separate the WBF signal from various QCD backgrounds, the following selection cuts are imposed: We require at least two hard jets with 
\beq
p_{T_j}\geq 20~\mathrm{GeV}\,,\;\quad\;
|y_j|\leq 4.5\,,
\eeq
where $p_{T_j}$ denotes the transverse component and $y_j$ the rapidity of the (massive) jet momentum which is reconstructed as the four-vector sum of massless partons of pseudo-rapidity $|\eta|<5$. The two reconstructed jets of highest transverse momentum are referred to as ``tagging jets''. We impose a large rapidity separation between the two tagging jets,
\beq
\Delta y_{jj} = |y_{j_1}-y_{j_2}|>4\,,
\eeq
and require that they be located in opposite hemispheres of the detector,
\beq
y_{j_1}\times y_{j_2}<0\,,
\eeq
with an invariant mass
\beq
M_{jj}>600~\mathrm{GeV}\,.
\eeq
For the charged leptons we request
\beq
p_{T_\ell}\geq 20~\mathrm{GeV}\,,\;\quad\;
|y_\ell|\leq 2.5\,,
\eeq
\beq
\Delta R_{j\ell}\geq 0.4\,,\;\quad\;
\Delta R_{\ell\ell}\geq 0.1\,,
\eeq
where $\Delta R_{j\ell}$ and $\Delta R_{\ell\ell}$ denote the jet-lepton and lepton-lepton separation in the rapidity-azimuthal angle plane. In addition, the charged leptons are required to fall between the two tagging jets in rapidity,
\beq
y_{j,min}<y_{\ell}<y_{j,max}\,.
\eeq

\begin{figure}
\begin{center}
  \includegraphics[width=0.4\textwidth,clip]{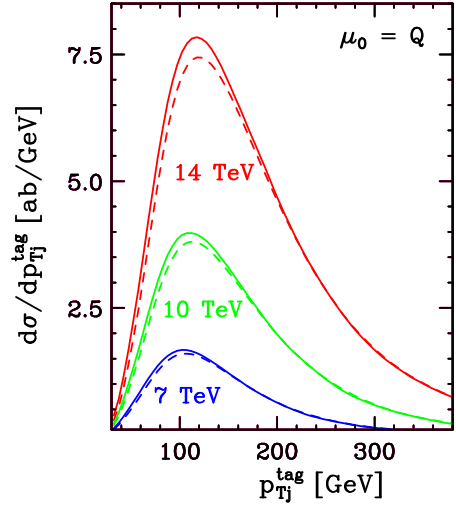}
  \caption{Transverse momentum distribution of the tagging jet with the highest $p_T$ in $pp\to e^+ \nu_e \mu^+\nu_{\mu} jj$ via WBF at LO (dashed lines) and NLO (solid lines) at the LHC for three different center-of-mass energies. }
  \label{fig:ptj-energy}
\end{center}
\end{figure}

In order to estimate the impact of NLO-QCD corrections on various kinematic distributions, we define the dynamical $K$ factor as
\beq
K(x) = \frac{d\sigma_\mr{NLO}/dx}{d\sigma_\mr{LO}/dx}\,.
\label{eq:kfac}
\eeq
Figure~\ref{fig:ptj} 
shows the transverse momentum distributions of the tagging jet with the highest $p_T$ in $pp\to e^+ \nu_e \mu^+\nu_{\mu} jj$ via WBF together with their $K$ factors for different choices of the factorization and renormalization scales, $\mu_F$ and $\mu_R$, which are taken as multiples of the scale parameter $\mu_0$, 
\beq
\mu_F = \xi \mu_0\,,\quad
\mu_R = \xi \mu_0\,.
\eeq
Results are shown for $\mu_0 = m_W$ and $\mu_0 = Q$, where $Q$ denotes the momentum transfer between an incoming and an outgoing parton along a fermion line. For both settings, we vary the scales in the range $\mu_0/2$ to $2\mu_0$. While the LO results are rather sensitive to $\mu_F$, the NLO curves barely depend on the scale choice in the considered range of $\xi$. In particular for $\mu_0=m_W$, the shape of $d\sigma/dp_{Tj}^{tag}$ changes noticeably when going from LO to NLO, as illustrated by the corresponding $K$ factors. Choosing $\mu_F=\mu_R=Q$ thus seems to be more suitable than $\mu_F=\mu_R=m_W$, should LO results be used to approximate jet distributions in WBF reactions. 
Figure~\ref{fig:ptj-energy} 
illustrates, how  $d\sigma/dp_{Tj}^{tag}$ changes, when the c.m.s.~energy is varied in the range of $\sqrt{S}=7$~TeV~to~14~TeV for $\mu_F=\mu_R=Q$. 

\begin{figure}
\begin{center}
\includegraphics[width=0.4\textwidth,clip]{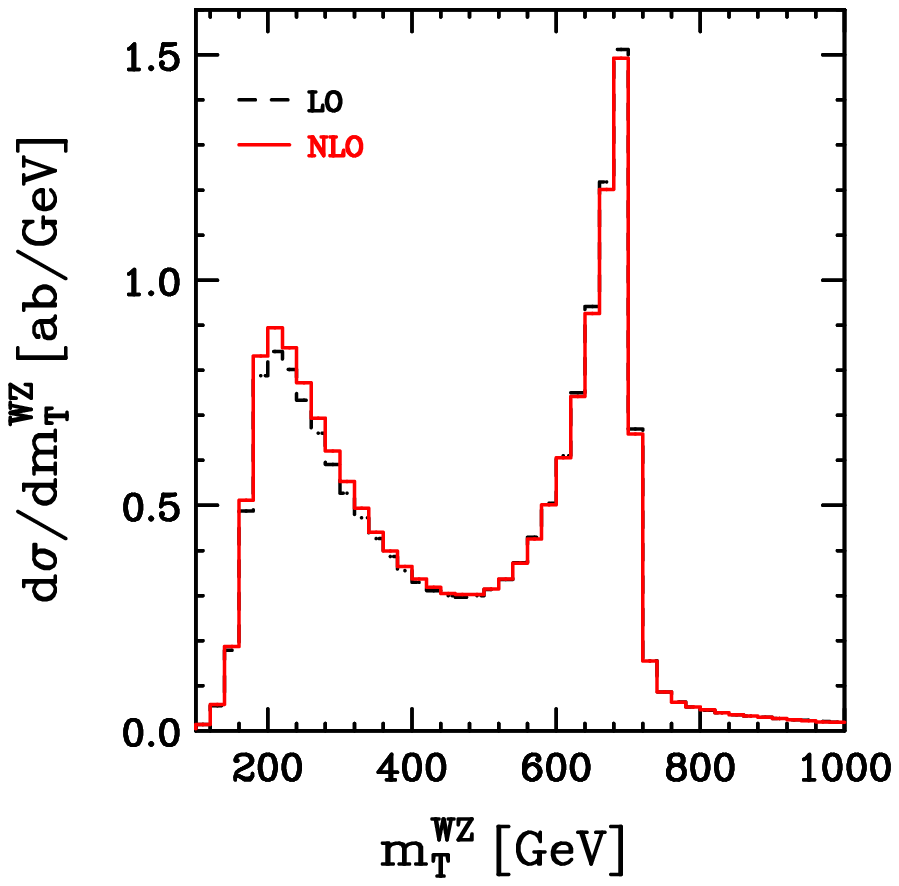}
\hs{1cm}
\includegraphics[width=0.4\textwidth,clip]{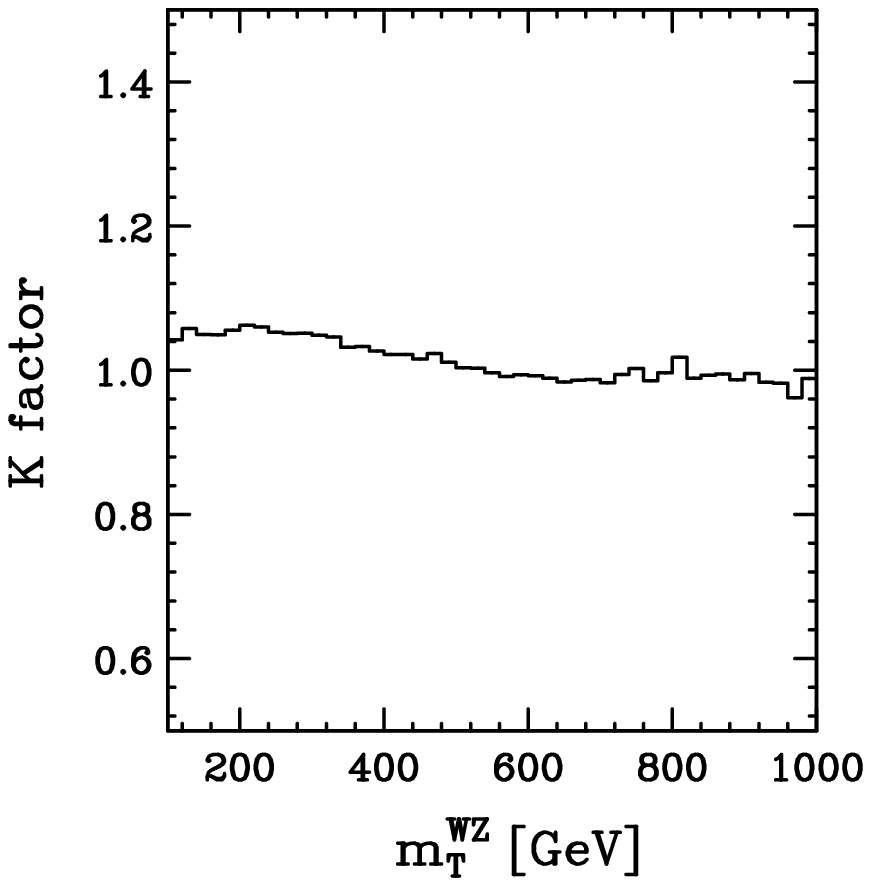}
\caption{\label{fig:mtnlo}
LO (black dashed line) and NLO (red solid line) distribution of the transverse cluster mass of the  $W^+Z$ system in a Warped Higgsless scenario (left) and differential $K$~factor (right). Scales are set to 
$\mu_R=\mu_F=Q$.}
\end{center}
\end{figure}

In Fig.~\ref{fig:mtnlo}  
we show the transverse cluster mass of the  decay-lepton system in $pp\to  e^+ \nu_e \mu^+\mu^- jj$ together with the differential $K$~factor in a representative Warped Higgsless scenario. As in the SM, NLO-QCD~corrections are small, but give rise to noticeable shape distortions. 

%%%%%%%%%%%%%%%%%%%%%%%%%%%%%%%%%%%%%%%%%%%%%%%%%%%%%%%%%%%%%%%%%%%
%
\section{Summary and Conclusions}
In this contribution, we have reviewed NLO-QCD calculations for weak boson scattering 
processes at the LHC within the SM and representative models of new physics. NLO-QCD corrections to total cross sections within WBF-specific selection cuts are moderate for all production modes. However, the shape of some distributions can change substantially beyond LO, in particular if a fixed factorization scale is used. 

%
%%%%%%%%%%%%%%%%%%%%%%%%%%%%%%%%%%%%%%%%%%%%%%%%%%%%%%%%%%%%%%%%%%%
%
\section*{Acknowledgments}
This work has been supported by the Initiative and Networking Fund of the
Helmholtz Association, contract HA-101 ("Physics at the Terascale").

%%%%%%%%%%%%%%%%%%%%%%%%%%%%%%%%%%%%%%%%%%%%%%%%%%%%%%%%%%%%%%%%%%%
%

\end{document}